\documentclass[apj]{emulateapj}
\usepackage{natbib}
\begin{document}

\title{Weak lensing mass reconstruction of the interacting cluster 1E0657-558:
Direct evidence for the existence of dark matter \footnote{Based on observations
made with ESO Telescopes at the Paranal Observatories under program IDs 60.A-9203
and 64.O-0332}}

\author{Douglas Clowe\altaffilmark{1}} 
\affil{Institut f\"ur Astrophysik und Extraterrestrische Forschung der
Universit\"at Bonn, Auf dem H\"ugel 71, 53121 Bonn, Germany}
\email{dclowe@as.arizona.edu}
\altaffiltext{2}{currently at Steward Observatory, University of Arizona}
\author{Anthony Gonzalez}
\affil{Department of Astronomy, University of Florida, 211 Bryant Space Science
Center, Gainesville, FL 32611-2055}
\and
\author{Maxim Markevitch}
\affil{Harvard-Smithsonian Center for Astrophysics, 60 Garden St., Cambridge, MA 02138}

\begin{abstract}
We present a weak lensing mass reconstruction of the interacting cluster 
1E0657$-$558 in which we detect both the main cluster and a sub-cluster. 
The sub-cluster is identified as a smaller cluster which has just undergone
initial in-fall and pass-through of the primary cluster, and has been previously
identified in both optical surveys and X-ray studies.  The X-ray gas has been
separated from the galaxies by ram-pressure stripping during the pass-through.
The detected mass peak is located between the X-ray peak and galaxy 
concentration, although the position is consistent with the galaxy centroid 
within the errors of the mass reconstruction.  We find that the mass peak for the
main cluster is in good spatial agreement with the cluster galaxies and offset from
the X-ray halo at $3.4\sigma$ significance, and determine that the mass-to-light 
ratios of the two components are consistent with those of relaxed clusters.  
The observed offsets of the lensing mass peaks from the peaks of the dominant
visible mass component (the X-ray gas) directly demonstrate the presence, and
dominance, of dark matter in this cluster.  This proof of the dark matter existence 
holds true even under the assumption of modified Newtonian gravity (MOND); from 
the observed gravitational shear to optical light ratios and mass peak -- X-ray gas 
offsets, the dark matter component in a MOND regime has a total mass which is at 
least equal to the baryonic mass of the system.
\end{abstract}

\section{Introduction}
It has been long established that the velocity dispersions and X-ray gas
temperatures of clusters of galaxies are too high to be explained solely by
the amount of visible matter in the clusters using
a physical model with Newtonian gravity and general relativity.  This observation led
to the introduction of a ``dark matter'' component of the mass which interacts
with normal matter and light only via gravity.  Recent observations of
clusters suggest that the mass is made of $\sim 1\%$ of baryons observable in
optical and infrared data, $\sim 11\%$ of baryons observable in X-ray data
(e.g.~\citealt{AL02.1}), and the remaining $\sim 88\%$ in the dark matter component.

An alternative explanation has been that the gravitational force only follows
the Newtonian $r^{-2}$ law at the level of the force observed in the solar
system, and that at smaller values the decline with distance is less 
\citep{MI83.1}.  This idea of modified gravity (MOND) has been 
used to reproduce the observed rotation velocities of spiral galaxies without
inclusion of any dark matter (e.g.~\citealt{MC98.1}), and could also explain 
the observed velocity dispersions and X-ray temperatures of clusters without 
needing any additional matter beyond the observed baryons and a small neutrino 
mass \citep{SA03.1}.  Several arguments against MOND have
been made using gravitational lensing observations of galaxies and clusters
\citep{GA02.1,HO02.1,MO01.1},
but because of the lack of a general relativistic theory giving the
strength of the interaction between light and gravity in the MOND regime
these observations can be explained by alterations in the MOND formalism
\citep{SA02.1}.

A definitive test of MOND, however, can be made with interacting clusters
of galaxies.  In the standard CDM paradigm, during the initial pass-through
the dark matter particles and the galaxies are effectively collisionless
while the X-ray halo is affected by ram pressure.  As a result, one
expects the galaxies and dark matter halo to remain spatially coincident
following the interaction, while the X-ray halo is displaced toward
the center of mass of the combined system (e.g.~\citealt{TO03.1}).  In the
CDM paradigm the mass of the X-ray halo is a small component of the total mass,
and therefore the mass maps created from weak lensing should have the primary
mass peaks in good spatial agreement with the galaxies.
In a MOND regime, however, the X-ray gas is the dominant component of the total
mass.  The weak lensing mass reconstruction would therefore detect a primary mass
peak coincident with the gas, which is spatially offset from the galaxy
distribution.

The $z = 0.296$ interacting cluster 1E0657-558 provides the ideal case in which to
test this theory. First discovered by \citet{TU95.1}, subsequent
analysis of {\sl ROSAT} HRI data revealed that the system is comprised
of two merging sub-clusters \citep{TU98.1}. More recent {\sl Chandra}
and spectroscopic observations further indicate that this merger is nearly
in the plane of the sky \citep{BA02.1,MA02.1},
with the lower mass sub-cluster having recently exited the core of the main
cluster \citep{MA02.1}.  

The {\sl Chandra} observations by \citet{MA02.1,MA03.2} have been
particularly valuable in elucidating the dynamical state and geometry of
this unique system. These data reveal the presence of a prominent bow shock
leading the lower mass sub-cluster ($T \sim 6$ keV), which is exiting the core
of the main cluster ($T \sim 14$ keV) with a relative velocity of 4500 km 
s$^{-1}$, determined from the gas density jump at the bow shock.  Coupled with 
the current 0.66 Mpc separation between the two 
components, this velocity requires that the closest approach of the two 
components occurred 0.1-0.2 Gyrs ago.  The merger is constrained to be nearly in
the plane of the sky by the sharpness of the shock front, a
result consistent with the small line-of-sight component of
the sub-cluster velocity derived from the spectroscopic data
by \citet{BA02.1}.  Finally, a 
comparison of the {\sl Chandra} data with optical imaging reveals that the 
X-ray gas associated with the bullet trails the galaxy distribution.  This 
latter result, coupled with the simple geometry of the system, enables the 
definitive test of MOND that we describe below. In an accompanying paper, 
\citet{MA03.1} utilize the combination of {\sl Chandra} and weak 
lensing data to also constrain the collisional cross-section for 
self-interacting dark matter.

This is not the only system which is known to have a spatial offset between
the galaxies and X-ray gas in a subcomponent.  The high-redshift cluster
MS1054-0321 has a double-peaked X-ray halo in which the western peak is offset
from the nearby galaxy overdensity \citep{JE01.1}.  Unlike 1E0657-558, however,
no shock front is observed in the X-ray data, and as a result the relative velocities 
and geometry of the merging components are unknown.  Further, while a weak lensing
mass peak has been measured near the galaxy overdensity \citep{HO00.1,CL00.1},
the uncertainty in the position of this mass peak is quite large \citep{MA02.2}.

In this paper we use $B$ and $I$ images taken with the FORS1 instrument in
direct imaging mode on the VLT1 8-m telescope during 1998 and 2000,
which were obtained from the ESO archive.  These include the images used in
\citet{BA02.1}, but we have independently created the final images from the raw
data.  In section 2 we present the weak lensing analysis of the image and
discuss the significance and uncertainties in the positions of the detected
mass peaks.  We analyze the photometry in section 3 and give mass-to-light
ratios for the detected mass peaks.  Discussion of the results and our
conclusions are presented in section 4.  Throughout this paper we assume
an $\Omega_\mathrm{m} = 0.3$, $\Omega_{\Lambda} = 0.7$, $H_0 = 70$ km/s/Mpc
universe unless stated otherwise.

\section{Weak Lensing Analysis}
Weak gravitational lensing is a method which can be used to measure the surface
mass in a region by utilizing the fact that the path of a light bundle passing
a gravitational potential will be bent by the potential.  As a result,
images of background galaxies which are near a massive structure, such as a
cluster of galaxies, are deflected away from the structure, enlarged while
preserving the surface brightness, and distorted such that they are stretched
tangentially to the center of the potential.  This third effect, known as
gravitational shear ($\gamma $), causes the
background galaxies' ellipticities to deviate from an isotropic distribution,
and the magnitude and direction of these deviations is used to measure the
mass of the structure(s) causing the lensing.  This technique of measuring
the mass does not make any assumptions about the dynamical state of the mass,
and is therefore one of the few methods which can be used to measure the
mass of a dynamically disturbed system.

\begin{figure*}
\plotone{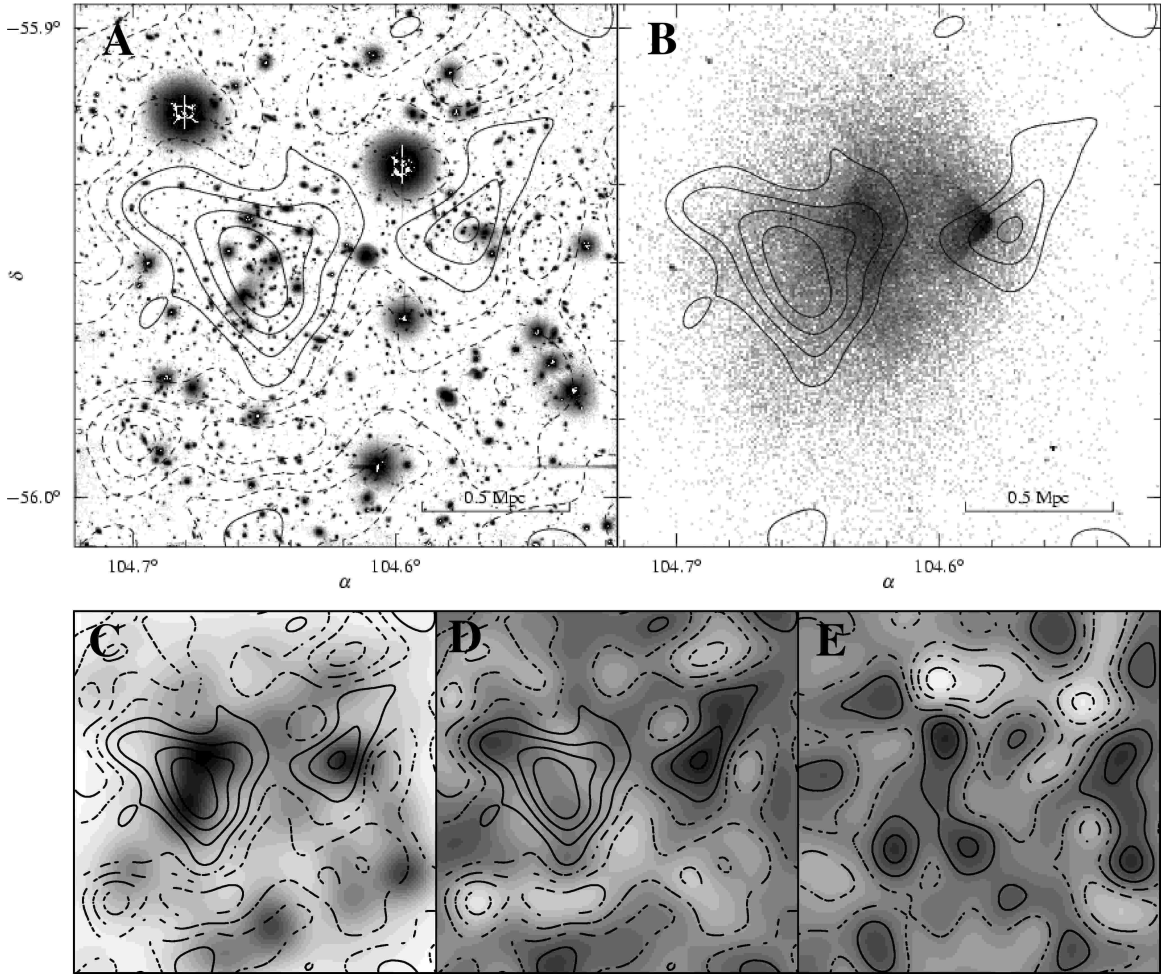}
\caption{A) Shown in greyscale is the $I$-band VLT image used to measure
  the galaxy shapes for the background galaxies.  Overlayed in black contours
  is the weak lensing mass reconstruction with solid contours for positive
  mass, dashed contours for negative mass, and the dash-dotted contour for the
  zero-mass level, which is set such that the mean mass at the edge of the
  image is zero.  Each contour represents a change in the surface mass density
  of $2.8\times 10^8 M_\odot/\mathrm{kpc}^2$.  B) Shown in greyscale is the 
  Chandra X-ray image from \citet{MA02.1} with the same
  weak lensing contours as in panel A.  C) Shown in greyscale is the
  luminosity distribution of galaxies with the same B-I colors as the primary
  cluster's red sequence.  Overlayed are the same mass contours as
  in panel A. D) Shown in greyscale is the mass reconstruction of the
  field after subtraction of the best-fit King shear profile for the primary cluster.
  Overlayed in are the same mass contours as in panel A.  E) Shown in
  greyscale, with the same color stretch as in panel D, is the mass reconstruction
  of the field after the background galaxies have been rotated by 45 degrees.
  This provides a good indication of the level of the noise in the reconstruction.
  The contours for the noise are drawn at the same values of $\kappa$ as
  for the mass reconstruction in panel A.}
\label{fig1}
\end{figure*}

The first step in the weak lensing analysis is to detect background galaxies,
measure their ellipticities, and correct the ellipticities for smearing due
to the point spread function.  We used the prescription given in \citet{CL02.1}
for performing this step, in which the objects are detected
and have their photometry measured using SExtractor \citep{BE96.1}, their shapes
measured using the IMCAT software package 
(http://www.ifa.hawaii.edu/\~kaiser/imcat), and the PSF smearing
correction performed using the KSB technique \citep{KA95.1}.  Background galaxies
were selected using the criteria of having $I > 20$, $B-I < 3.2$, having
a detection significance in $I$ greater than 11, and having a $50\%$ encircled
light radius larger than that of stars.  This selection resulted in a catalog
with a density of 12.3 galaxies/arcmin$^2$ over a box $6\farcm 7$ on a side, 
which is complete, as measured by the departure of the number counts from a 
power law, to $I\sim 24.5$, with the faintest galaxy having $I=25.97$.

The next step in weak lensing analysis is to convert the measured shear into
a measurement for the convergence $\kappa$, which is related to the surface
density of the lens $\Sigma$ via
\begin{equation}
\kappa = {\Sigma \over \Sigma _{\mathrm{crit}}}.
\end{equation}
where $\Sigma _{\mathrm{crit}}$ is a scaling factor:
\begin{equation}
\Sigma _{\mathrm{crit}} = {c^2 \over 4 \pi  G}{D_{\mathrm{s}} \over 
D_{\mathrm{l}} D_{\mathrm{ls}}}
\end{equation}
where $D_{\mathrm{s}}$ is the angular distance to the source (background) 
galaxy, $D_{\mathrm{l}}$ is the angular distance to the lens (cluster), 
and $D_{\mathrm{ls}}$ is the angular distance from the lens to the source 
galaxy.  Using the same magnitude and color selections on the HDF-S 
photometric redshift catalog from \citet{FO99.1} as were used to
create the background galaxy catalog results in a mean lensing redshift of
$z_\mathrm{bg} = 0.85$, and a $\Sigma _{\mathrm{crit}} = 3.1\times 10^9
M_\odot/\mathrm{kpc}^2$.

Shown in Fig.~\ref{fig1} in solid dark contours is a map of $\kappa$
for this field created by using the KS93 algorithm \citep{KA93.1} which uses the fact
that both the shear and the convergence are combinations of various second
derivatives of the surface potential, and therefore the Fourier transform
of the shear can be converted into the Fourier transform of $\kappa$ by
the multiplication of the appropriate wave numbers.  Because we are
reconstructing a small field around a massive cluster, however, we actually
measure the reduced shear $g = \gamma /(1-\kappa )$ from the background
galaxy ellipticities.  Therefore we must perform an iterative solution
to the KS93 algorithm in which an initial $\kappa$ map is assumed (in this
case $\kappa = 0$ everywhere), $g$ corrected with this map to $\gamma $,
which is then transformed to a $\kappa $ map, which is then used in turn
to correct $g$, etc \citep{SE95.1}.  This technique typically converges in a few
iterations (in this case 6), and gives an measurement of $\kappa $ in the
field relative to the level of $\kappa $ at the edge of the image, which
is unknown.

As can be seen in Fig.~\ref{fig1}, two distinct mass peaks are found in the field,
each of which is spatially coincident with an overdensity of galaxies.
Spectra for galaxies in both structures have been published in \citet{BA02.1} and 
the two
groups have the same redshift.  The peaks have significances of 6.4$\sigma$
for the larger eastern peak (hereafter referred to as ``the cluster'') and
3.0$\sigma$ for the smaller western peak (hereafter referred to as 
``the subclump'').  The significances were measured by convolving the
mass maps with Mexican-hat filters and comparing the filtered value at the
peak position with those of randomizations of the mass maps.  
The randomizations were performed by first subtracting a smoothed value of
the shear (smoothed using a $22\farcs 4$ Gaussian weighted average of the
surrounding galaxy ellipticities) from the galaxy shear estimates to obtain 
an estimate of the intrinsic ellipticity of the galaxies, then applying a 
random spin to the orientation of each background galaxy while preserving their
positions and intrinsic ellipticities, and creating mass maps from the
catalogs.  

An X-ray luminosity map from Chandra data
\citep{MA02.1} is overlayed in grey contours in Fig.~\ref{fig1}.  As can be
seen, both peaks are also visible in the X-ray data, but are offset in
position from both the galaxies and the mass peaks.  From the shape,
strength, and location of the shock visible in the X-ray peak for the
subclump,
\citet{MA02.1} have concluded that this system has just undergone initial
infall and pass-through, and the two clusters are now moving away from one
another.  The separation between the galaxies, which are effectively 
collisionless particles in such a pass-through event, and the X-ray gas is
a result of the ram pressure of the interacting gas halos slowing down the
X-ray halos during the interaction.  As a result, a separation between
the mass peak and the X-ray peak and an agreement in positions between
the mass peak and galaxy overdensity would suggest that the dark matter component
of the cluster must be relatively collisionless, as compared to the X-ray
emitting baryonic gas.  

\begin{figure}
\plotone{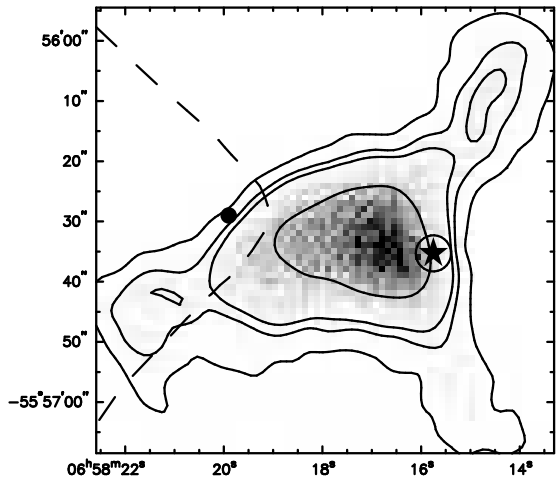}
\caption{Shown above is the error map for the centroid of the weak lensing
  mass peak associated with the subclump, generated using 10,000 bootstrap
  resamplings of the background galaxy catalog.  The thick black contours indicate the
  regions containing 68\%, 90\%, 95\%, and 99\% of the centroid positions
  after smoothing with a $2\arcsec $ FWHM Gaussian kernel.  The solid
  circle shows the position of the X-ray peak associated with the subclump
  and the dash contour shows the boundary of the gas associated with the
  subclump.  The solid star shows the centroid of the galaxies in the subclump
  with the encompassing circle showing the $1\sigma$ error contour of the centroid.}
\label{fig2}
\end{figure}

In order to place limits on the collisional cross-section of dark matter
from the displacement of the mass peak from the X-ray peak,
we calculated the error on the centroid determination of the subclump by
performing mass reconstructions on 10,000 bootstrap-resampled catalogs of
the background galaxies.  For each reconstruction the $\kappa$ map was 
convolved with a Mexican-hat filter to detect the nearest peak to the position
of the subclump and measure its significance, imposing a minimum significance
of $1\sigma$.  The resulting distribution of positions is plotted in Fig.~\ref{fig2}.
As approximately $2.5\%$ of the reconstructions should have the mass
peak associated with the subclump at less than $1\sigma $ significance,
we eliminated the 250 most distant peaks from the position of the subclump in
the data.  The remaining peaks have a rms positional offset of $12\farcs1$.  
The separation between the mass peak and X-ray peak in the data is
$22\farcs6$, which is significant at a 1.9$\sigma$ confidence level.  This 1-D
error analysis, however, is at some level incorrect as the distribution of the
peak positions is not a circular Gaussian and the
resampled peak distribution has a larger rms errors in right ascension than in
declination.  To measure the significance in the 2-D peak position distribution we
binned the data into $1\arcsec \times 1\arcsec$ bins and drew contours of
decreasing number of peaks until the contour intersected the position of the
X-ray peak.  Located inside this contour were $95.5\%$ of the resampled peaks.
We discuss constraints that this system gives on
the collisional cross-section of dark matter in a related paper 
\citep{MA03.1}. 

The X-ray gas of the cluster is also offset from the cluster galaxies and
associated dark matter peak.  The dark matter peak is in good spatial agreement
with the cluster galaxies, and the difference in the shape of the dark matter
peak relative to the galaxy luminosity distribution seen in Fig.~\ref{fig1}
is consistent with being caused by the noise in the mass reconstruction
\citep{CL00.1}.  Using the same bootstrap-resampled catalogs described above
and looking for the nearest peak to the position of the cluster gives the 
significance of the offset between the X-ray gas and dark matter to be 
$\sim 3.4\sigma$.  The offset gas, however, is a combination of the gas from the cluster and
gas stripped from the outskirts of the subclump, and therefore requires more
complicated physics to interpret.  

\begin{figure}
\plotone{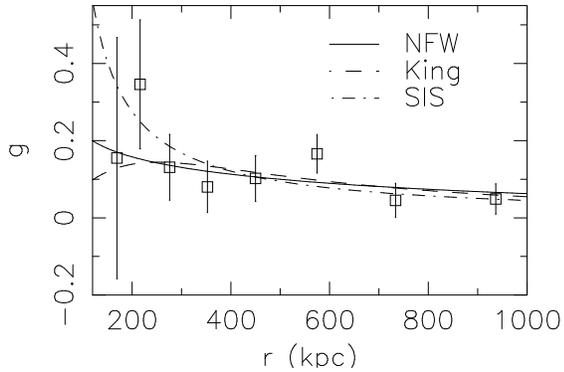}
\caption{Plotted above is the reduced shear profile for the main cluster in
the 1E0657$-$558 system.  Also shown are the reduced shear profiles for the
best-fit NFW (solid line), King (dashed line), and SIS (dash-dotted line) 
models.}
\label{fig3}
\end{figure}

Because the KS93 mass reconstruction can only measure the mass relative to
the mean mass at the edge of the field and that the images are smaller than
the expected dynamical size of the cluster ($6\farcm 7 = 1770 \mathrm{kpc}$),
one cannot measure the mass of the cluster reliably with the mass 
reconstruction in Fig.~\ref{fig1}.  Instead, we have measured the mass of the cluster
using radial shear profile fitting in which one assumes a surface mass model
for the cluster, converts this into a $\kappa $ profile, and then into a
profile for the reduced shear which is compared to the azimuthally averaged
shear profile from the data, as shown in Fig.~\ref{fig3}.  We tried fitting a singular isothermal sphere, a
NFW model \citep{NA96.1}, and a King model to the data and found that the King model
was marginally preferred over the NFW model, as measured
by the $\delta \chi ^2$ between the model reduced shear profile and the data. 
Using an F-test \citep{BE92.1} to compare the 1-parameter SIS to the 2-parameter
NFW and King models resulted in both the NFW and King models being preferred
to an SIS at $91\% $ confidence.  We excluded a $1\arcmin$ diameter region 
around the
subclump from the shear profile in order to minimize any contamination of the
profile from the subclump.  Even with this excluded area, however, the fit
will still include the mass of the subclump in the total mass of the cluster
for radii larger than the subclump-cluster separation which will have the
effects of overestimating the total mass of just the cluster itself as well
as underestimating the concentration of the cluster.  It should also be noted
that at small smoothing lengths, the mass reconstruction of the field shows
two mass peaks for the main cluster, and thus the fact that the King
core-model profile is the preferred mass profile may be due to the blending of
two peaks in the radial profile rather than a core in a single peak.

The King model has a mass density profile
\begin{equation}
\rho (r) = {\rho _0 \over [1 + {r^2 \over r_\mathrm{c}^2}]^{3\over 2}}
\end{equation}
which integrates to have a surface density profile
\begin{equation}
\Sigma (x) = {2 \rho _0 r_\mathrm{c} \over 1 + {x^2 \over r_\mathrm{c}^2}}
\end{equation}
where $\rho _0$, the central mass density, and $r_\mathrm{c}$, the core radius,
are the fitting parameters, $r$ is the 3-D radius, and $x$ is the 2-D 
projected radius.  The integrated NFW profile can be found in \citet{BA96.1}.
The best-fit parameters were $\rho _0 = 3.85\times 10^6 M_\odot/\mathrm{kpc}^3,
r_\mathrm{c} = 214 \mathrm{kpc}$ for the King model and $r_{200} = 2250 \mathrm{kpc}$, 
$c = 3.0$ for the NFW profile.  Both models have the two
parameters degenerate in the fits with poor constraints on both $c$ and
$r_\mathrm{c}$.  The significances of the fits, as measured by the 
$\delta \chi ^2$ between the model fit and a zero mass model fit, is
6.48 for the King model and 6.37 for the NFW model.  The King model has
surface mass measurements of $9.5\pm 1.5 \times 10^{13} M_\odot $,
$2.0\pm 0.3 \times 10^{14} M_\odot$, and $4.4\pm 0.7 \times 10^{14} M_\odot$
for 150, 250, and 500 kpc respectively.  The NFW model has surface mass
measurements of $1.02\pm 0.16 \times 10^{14} M_\odot$, $2.1\pm 0.3 \times 10^{14}
M_\odot$, and $5.3\pm 0.8 \times 10^{14} M_\odot$ for the same radii.  These
masses are in good agreement with the velocity dispersion for early-type galaxies
given by \citet{BA02.1}.

\begin{figure}
\plotone{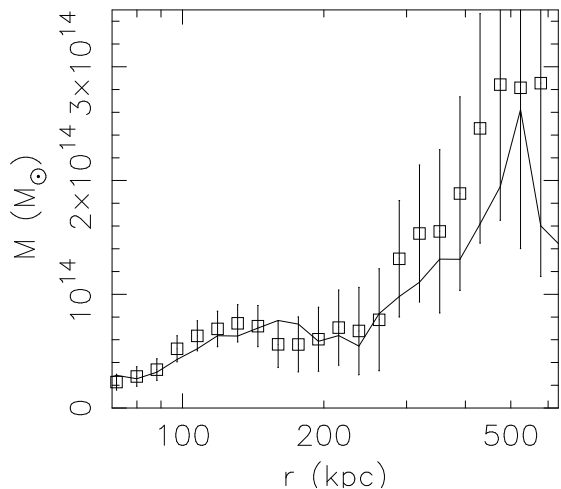}
\caption{Plotted above is the minimum surface mass profile for the subclump
generated using aperture densitometry centered on the centroid of the galaxy
distribution.  The solid line is the mass profile when centered on the mass
peak in the 2-D mass reconstruction.  The error-bars for aperture densitometry
are correlated such that every data point has knowledge of the values of the
points at larger radius.}
\label{fig4}
\end{figure}

In order to measure the mass of the subclump, we first had to remove the mass
of the main cluster which would otherwise provide a large positive bias to
the mass measurements.  This was accomplished by subtracting the reduced shear
profile of the best-fit King model for the cluster from the background galaxy 
catalog.  A mass reconstruction of this catalog shows that the main cluster
has been effectively removed from the lensing signal, as can be seen in 
Fig.~\ref{fig1}.  We fitted the three mass model profiles to the
subclump shear profile, and while the King model was the preferred model,
its $\chi ^2$ indicated that it was not a good fit to the data.
Instead, we have measured the mass of the subclump using aperture
densitometry \citep{CL00.1,FA94.1}, which measures the mean surface density inside a 
cylinder of a given radius minus the mean surface density in an annular region.

Using a 150 kpc radius for the disk and an annular region with radial extent
of 640 kpc to 706 kpc resulted in a mass measurement of $7.3\pm 2.1 \times 
10^{13} M_\odot $ when centered on the middle of the subclump mass peak seen
in Fig.~\ref{fig1} and $6.6\pm 1.9 \times 10^{13} M_\odot $ when centered on the
centroid of the red subclump galaxies.  The choice of a 150 kpc radius for the
mass measurement was made for two reasons:  First, the mass profile for the 
subclump, shown in Fig.~\ref{fig4}, shows evidence for a plateau in mass associated
with the subclump between 150 and 250 kpc.  This might indicate that the dark matter
at larger radii has been tidally stripped from the core during the interaction.
Secondly, the noise level in the aperture densitometry increases dramatically 
with radius, so that the signal-to-noise of the mass measurement goes from more than 
$3\sigma$ at $100 \mathrm{kpc} \la r \la 150 \mathrm{kpc}$ to less than $1.5\sigma$
at $r\ge 300 \mathrm{kpc}$.

The observed shear and derived mass of the subclump are significantly higher
than could be produced by an isothermal sphere with a 212 km/s velocity dispersion,
as measured by \citet{BA02.1}.  This velocity dispersion, however, is measured
from only 7 galaxies and could be biased low by their method for distinguishing a
cluster galaxy from a subclump galaxy.  Further, the conversion of a velocity
dispersion to mass measurement requires the assumption of virial equilibrium,
which is unlikely to apply to the subclump.  The weak lensing mass is in good
agreement with the 7 keV X-ray temperature for the cold gas blob, under the 
assumption that this was the temperature of the subclump prior to the interaction.

\section{Photometry}
\subsection{Centroid of the Galaxy Distribution}

For the standard CDM paradigm we expect the dark matter distribution
to be coincident with the galaxy distribution if the dark matter
particles are collisionless. To test this expectation we employ
adaptive kernel smoothing to determine the centroid of the galaxy
distribution associated with the lower mass sub-cluster. The method is
similar to that described in \citet{GO02.1}. An Epanechnikov
kernel with $h=30\arcsec$ is used for the adaptive smoothing and the
highest density peak within a $50\arcsec$ radius is identified as the
sub-cluster. We restrict the input catalog to objects with m$_I$=18-25,
$B-I$ color within 0.5 magnitudes of the brightest cluster galaxy
($B-I=3.9$), and SExtractor stellarity indices less than 0.5 in both
bands. We compute the number-weighted and flux-weighted centroids,
finding that both weighting schemes yield indistinguishable
results. The uncertainty is calculated by recomputing the centroid for
1,000 bootstrap-resampled catalogs.

We find that the peak of the galaxy distribution is coincident with
the location of the brightest cluster galaxy. The number-weighted
centroid is $06:58;15.66,-55:56:35.3$ with a 1.9$\arcsec$
1$\sigma$ Gaussian uncertainty, while the flux-weighted centroid is
$06:58:15.75,-55:56:35.3$ with a 3.0$\arcsec$ 1$\sigma$
uncertainty. The net separation between the galaxy and weak lensing
centroids is $12\farcs 3$, which has a significance of $1\sigma$ in the 1-D
error analysis and $70\%$ in the 2-D error analysis described in the previous
section. 

\subsection{Luminosity}

The luminosity is determined using two approaches. We first derive the
luminosity from the galaxy catalog, which is the typical method used
in cluster $M/L$ measurements. We then directly compute the luminosity
by integrating the total flux within the aperture, which places a hard
upper limit on the luminosity.  For both techniques we correct the
derived luminosities for extinction \citep{SC98.1} and apply
$e+K$ corrections to convert to rest-frame passbands. The applied
corrections are based upon the Bruzual \& Charlot evolutionary models
for a passively evolving elliptical galaxy
\citep{BR93.1,CH96.1}. The integrated $B-I$ color is
consistent with the prediction for passive evolution, indicating that
this approximation is reasonable. Absolute magnitudes are converted to
units of solar luminosity using the solar absolute magnitudes given in
Allen's Astrophysical Quantities \citep{AL00.1}.

For the catalog approach, we first cull the input photometric catalog to
minimize contamination from stars and foreground galaxies.  We exclude
all objects that are brighter than the BCG, more than 1 magnitude
redder than the BCG (i.e. $B-I<4.9$), or have stellarity index $>$0.8
and $\mathrm{m}_I<20$. Two foreground spirals are also
removed from the input catalog.  The flux from the remaining galaxies
is then summed, with the resulting luminosity shown in Table
\ref{table1}. We caution that there are two caveats with this
approach. First, we lack sufficient spatial coverage to employ
background subtraction. The impact of background contamination is
expected to be minor because of the large density contrast within our
physical apertures, but such contamination will yield a positive bias
in the derived luminosity.  This effect will be greatest for the largest
apertures.  Second, incompleteness at the faint end of
the luminosity function yields a negative bias in the derived
luminosity. Because our data is complete to roughly 5 (6) magnitudes
below $L_*$ in $I$ ($B$), the magnitude of this effect should be 4.5
(2.5)\% for a faint end slope $\alpha=-1.3$.

Directly integrating the flux within the apertures provides a useful
cross-check on the above technique. For this approach we only mask
stars with stellarity index $>$0.8 and m$_I<18$. We then
integrate the flux within the aperture, using two additional apertures
located 2$\arcmin$ north and south of the cluster to quantify the
background sky level.  The sky level in these apertures were computed
after masking detected objects in the region, and thus the detected
background level does not include flux associated with resolved galaxies
(except that scattered onto the extreme wings of the PSF).  Background
galaxy contamination is expected to yield a slight positive bias in
the derived luminosity, as with the catalog approach.
The results are shown in Table \ref{table1}. The
luminosities derived via this method typically agree the catalog results 
to within 15\%.

\subsection{Mass-to-Light Ratios}

We determine the mass-to-light ratios of both the main cluster and the
sub-cluster in rest-frame $B$ and $I$. If the sub-cluster has suffered
significant mass loss during passage through the core of the main
cluster, this should be reflected by a decreased mass-to-light
ratio. We find no evidence for such mass loss, with the $M/L$ ratios
for both components consistent with one another to within the
$1\sigma$ uncertainties. This result implies, under the assumption that
the initial $M/L$ ratio for the two structures were similar, that the dark matter
interaction cross-section must be small, a topic that is explored in
greater detail in an accompanying paper \citep{MA03.1}.  We
further note that the derived $M/L$ values are consistent with other
recent lensing-derived mass-to-light ratios.  \citet{DA00.1} for
instance finds $M/L_B=259\pm12$ for a sample of 40 low-redshift clusters.
The low dispersion of the $M/L$ ratios in the \citet{DA00.1} sample
also suggests that the assumption of similar intrinsic $M/L$ ratios for
the two components in this cluster is not unreasonable.

For these mass-to-light ratio calculations for the subclump we use an
aperture centered on the centroid of the galaxy light distribution,
assuming that the offset in the mass peak seen in Fig.~\ref{fig1} is a result of
noise in the shearfield.  If instead we use an aperture centered on the
observed mass peak for the subclump, the mass-to-light ratio of the subclump
increases by $\sim 10\%$.  Also, the mass estimates for the subclump were
created by subtracting the mean surface density in a $640-760$ kpc annulus
from the mean surface density within the 150 kpc disc, and since no similar
subtraction was performed on the light or the cluster mass, the mass-to-light
ratio of the subclump must be considered a minimum value in making comparisons
with the main cluster.

\begin{deluxetable*}{lcccccc}
  \tabletypesize{\small}
  \tablecaption{Mass-to-Light Ratios\label{table1}}
  \tablehead{\colhead{Region} & \colhead{$R$} & \colhead{$M$}  & \colhead{$L_B$}  &
    \colhead{$L_I$}  & \colhead{$M/L_B$} & \colhead{$M/L_I$} \\
    & (kpc) & ($10^{14} M_\odot$) & ($10^{11} L_\odot$) &
    ($10^{11} L_\odot$) &  &  \\}
  \startdata
  \multicolumn{7}{c}{Integrated Flux Technique} \\
  Sub-cluster & 150 & .66$\pm$.19 & 2.1 & 4.9 & 314$\pm$90 & 135$\pm$39   \\
  Main & 150 & .95$\pm$.15 & 3.5 & 7.8 & 271$\pm$43 & 122$\pm$19  \\
  Main & 250 & 2.0$\pm$0.3 & 8.5 & 15.8 & 235$\pm$35 & 127$\pm$19 \\
  Main & 500 & 4.4$\pm$0.7 &17.4 & 32.5 & 253$\pm$40 & 135$\pm$22 \\
  \multicolumn{7}{c}{Catalog Technique} \\ 
  Sub-cluster & 150 &       & 2.4 & 3.8 & 275$\pm$79 & 174$\pm$50  \\
  Main & 150 &             & 3.2 & 6.3 & 297$\pm$47 & 151$\pm$19  \\
  Main & 250 &             & 8.1 & 12.9 & 247$\pm$37 & 155$\pm$23 \\
  Main & 500 &             &21.4 & 28.8 & 206$\pm$32 & 152$\pm$24 \\
  \enddata
  \tablecomments{Due to the different method in which the main cluster and subclump
masses were measured, the subclump $M/L$ ratios must be considered a lower bound when
comparing with the main cluster.}
\end{deluxetable*}

\section{Discussion and Conclusions}
In a CDM universe, one would expect that the mass peaks for the cluster and
subclump would agree with the
centroid of the galaxy distributions, as both galaxies and dark matter
particles are collisionless in such an interaction \citep{TO03.1}.
One would also expect that the mass-to-light ratios would
decrease by $\sim 10-15\%$ as compared to relaxed systems due to the baryonic 
X-ray halo mass being removed from the structures.  Such a scenario is, 
within errors, in good agreement with the data.

In a purely baryonic MOND universe the X-ray and galaxy centroids would still
be separated as the galaxies are still collisionless particles in the
interaction.  However, because the X-ray halo is the dominant mass component of the
visible baryons in the cluster, in the absence of a dark mass component the vast
majority, $\sim 85-90\%$, of the mass of the subclump would be with the X-ray
gas.  Thus, any direct method to measure the mass of the system would detect
a higher mass about the stripped X-ray halo than around the galaxies.  This is
not what is observed in this system.  In order to quantify how much these observations
disagree with MOND, however, we first need to determine a method to measure the
masses of the clusters in a MOND cosmology.

Unfortunately, because there is not a derivation of MOND from general
relativity, there is not a definitive way to measure a mass with weak lensing from a
measured shear field.  If one assumes that the relation from general
relativity between the deflection of a photon and of a massive particle moving
at the speed of light by a static gravitational field is unchanged by MOND, 
then it can be shown that the shear field caused by a point mass is
\begin{equation}
\gamma (\theta ) = {\theta ^2_\mathrm{E} \over \theta _0 \theta ^2} \left[
{\theta \over 2} + \theta _0 - {\theta ^3 \over 2 (\theta + \theta
  _{\mathrm{out}})^2} \right],
\end{equation}
where $\theta $ is the distance from the point mass, $\theta _\mathrm{E}$ is
the Einstein radius for the lens, $\theta _0$ is the distance at which the
gravitational acceleration changes from Newtonian to MOND, and $\theta
_\mathrm{out}$ is the distance at which the gravitational acceleration changes
back to a $\theta^{-2}$ law \citep{MO01.1,HO02.1}.  As both $\theta _\mathrm{E}$ and
$\theta_0$ scale as the square root of the point mass, the resulting shear profile
scales linearly with the mass for $\theta \ll \theta _0$ and $\theta \gg
\theta _{\mathrm{out}}$, as the square root of the mass for $\theta _0 \ll
\theta \ll \theta _{\mathrm{out}}$, and somewhere between the two extremes
for the transitional regions $\theta \sim \theta_{0}$ and $\theta \sim \theta 
_{\mathrm{out}}$.

Calculating the expected shear profile for an extended source in the MOND
regime is complicated by the lack of a thin lens approximation, which is used
to simplify the equations with Newtonian gravity \citep{MO01.1}.  However, it is reasonable
to assume that the same general relation between the gravitational shear field
and overall mass of a halo exists as per the point mass relation.  As such,
the level of weak shear produced by a cluster of galaxies at a radii of a few
hundred kpc from the cluster core (which would be between $\theta _0$ and
$\theta _\mathrm{out}$ for $10^{13}$ to $10^{16} M_\odot$ clusters) should
scale with the mass of the cluster, probably somewhere between a linear scale
and a scale with the square root of the mass.  

From observations, we know that the shear fields produced by
individual galaxies \citep{HO03.1,MC02.1} are an order of magnitude lower than those produced by
galaxy groups \citep{HO01.1}, which are an order of magnitude lower than those produced by
poor clusters \citep{WI00.1}, which are in turn significantly lower than those produced by
rich clusters \citep{DA02.1,CL00.1,CL01.1,CL02.1}.  As the amount of visible baryons in these structures
scale in a similar manner, then from these observations we have support for the
above assumptions.  Thus, in a MOND universe, one should still observe a change in the 
shear field of a structure with a change in the mass of the structure.  

As a result, if the mass of
clusters of galaxies is limited solely to visible baryons, then by removing
the X-ray halo from the cluster one should reduce the gravitational shear
centered on the galaxies by at least a factor of three, if the shear scales as
the square root of the mass, and up to a factor of ten if the shear scales
linearly with the mass.  In this system, however, we find that the ratio of the 
gravitational shear to visible light for two components which have the X-ray halo
stripped from the galaxies is consistent with that found in normal clusters, which 
have the X-ray halo and galaxies spatially coincident.
This is inconsistent with the shear scaling as the square root of the mass
MOND model at roughly a $2\sigma$ confidence level and with the shear scaling linear with
the mass at roughly a $3\sigma$ confidence level.  

In order to reduce the inconsistency with the data to a $\sim 1\sigma$ confidence
level, one would need to add a non-luminous mass component to the clusters
which is equal to the mass of luminous matter for the shear scaling as square
root of mass case, and which is 2.5 times the mass of luminous matter for
shear scaling linearly with mass case.  This extra mass component would also
reduce the problem with the detected mass peak for the subclump being closer to the 
galaxies than the X-ray halo as the detected signal would be a blend of the two
components due to the required smoothing of the mass map.  

The more significant
offset between the cluster mass peak and X-ray halo would require a greater amount
of dark mass to explain if the two components were cleanly separated.  The X-ray
halo, however, is extended over the cluster galaxies which may indicate some fraction
of the X-ray gas has already been drawn back to the galaxy position.

Any dark mass component of the system must be
relatively collisionless, so it can undergo the pass-through without loss of
velocity or mass, and able to clump on scales smaller than 100 kpc (the
smallest aperture for which we can reliably measure the shear about the 
subclump).  Adopting big-bang nucleosynthesis limits on the mean baryonic mass
of the universe excludes most of this mass from being baryons in cold,
condensed structures.  The clumpiness limit excludes the matter from being
massive neutrinos with masses less than 4.5 eV \citep{SA03.1,TR79.1}.  Since
neutrinos more massive than 2.2 eV have been ruled out experimentally
\citep{BO02.1,LO01.1}, neutrinos thus cannot explain this mass.

\subsection{Summary}
We have shown above that the cluster 1E0657$-$558 has a lower-mass subclump
visible in X-ray and optical observations as well as in a weak lensing
mass reconstruction.  The X-ray and optically luminous components are spatially
separated at high significance, as one would expect for a system which
has just undergone an initial infall and transit of a larger mass system
\citep[e.g.][]{TO03.1}.  The observed mass peak in the weak lensing
reconstruction lies between X-ray and optical components, but is closer to, and
consistent with, the
optical component.  The centroid of the subclump mass peak has a fairly large
error resulting in the offset of mass peak from the centroid of the galaxy
distribution having a $\sim 70\%$ confidence level and the offset of the mass peak
from the X-ray peak having a $\sim 95\%$ confidence level.  

The primary cluster has also been detected in the weak lensing mass reconstruction, 
and has a mass peak which is spatially coincident with the cluster galaxies.  The
X-ray gas from the main cluster is offset from the mass peak at a $3.4\sigma$
significance.

We have also measured the mass-to-light ratio for the subclump at a 150 kpc radius
and for the main cluster at 150, 250, and 500 kpc radii.  We find that the
subclump has a mass-to-light ratio which is consistent with
the mass-to-light ratio of the main cluster, and that both are consistent with
mass-to-light ratios for relaxed clusters.  
The dominant source of error in the mass-to-light ratios and the mass--X-ray gas
offsets is the weak lensing mass reconstructions, which can be improved by 
obtaining shear
information on a wider field than the $7\arcmin \times 7\arcmin$ VLT field
and/or by obtaining deeper imaging on the same field with a smaller PSF in
order to greatly increase the number density of background galaxies usable for
the measurement of the shear field.

Finally, we have argued that even in a MOND universe, a significant fraction
of the original mass of the subclump must exist in the form of dark matter which,
furthermore, should be non-baryonic and non-neutrino.  The exact amount of extra 
mass cannot be calculated due to the lack of a MOND derivation from general 
relativity, but phenomenological arguments suggest that it is at least equal 
to the baryonic mass of the cluster.  While these observations cannot disprove 
MOND, or alternatively prove that gravity is Newtonian on small acceleration scales, 
they remove its primary motivation of avoiding the notion of dark matter.

\acknowledgements
We wish to thank Oliver Czoske and Alexey Vikhlinin for useful discussions.
This work was supported by the Deutsche Forschungsgemeinschaft under the project 
SCHN 342/3--1 (DC).  AHG is supported by a NSF Astronomy and Astrophysics
Postdoctoral Fellowship under award AST-0407485.  MM received support by 
NASA contract NAS8-39073, {\emph Chandra} grant GO2-3165X, and the
Smithsonian Institution.

\bibliographystyle{apj}
\bibliography{1e0657}

\end{document}